# Newly Developed Flexible Grid Trading Model Combined ANN and SSO algorithm

Wei-Chang Yeh [1,*], Yu-Hsin Hsieh [1], Chia-Ling Huang [2]

[1] Department of Industrial Engineering and Engineering Management, National Tsing Hua University, Hsinchu 300044, Taiwan; W.C. Yeh: yeh@ieee.org; Y.H. Hsieh: kypss93003@gmail.com

[2] Department of International Logistics and Transportation Management, Kainan University, Taoyuan 33857, Taiwan; clhuang@mail.knu.edu.tw

* Correspondence: yeh@ieee.org

**Abstract**: In modern society, the trading methods and strategies used in financial market have gradually changed from traditional on-site trading to electronic remote trading, and even online automatic trading performed by a pre-programmed computer programs because the continuous development of network and computer computing technology. The quantitative trading, which the main purpose is to automatically formulate people's investment decisions into a fixed and quantifiable operation logic that eliminates all emotional interference and the influence of subjective thoughts and applies this logic to financial market activities in order to obtain excess profits above average returns, has led a lot of attentions in financial market. The development of self-adjustment programming algorithms for automatically trading in financial market has transformed a top priority for academic research and financial practice. Thus, a new flexible grid trading model combined with the Simplified Swarm Optimization (SSO) algorithm for optimizing parameters for various market situations as input values and the fully connected neural network (FNN) and Long Short-Term Memory (LSTM) model for training a quantitative trading model to automatically calculate and adjust the optimal trading parameters for trading after inputting the existing market situation is developed and studied in this work. The proposed model provides a self-adjust model to reduce investors' effort in the trading market, obtains outperformed investment return rate and model robustness, and can properly control the balance between risk and return.

## 1. Introduction

The formulation of financial market trading methods and strategies has changed with the continuous development of network and computer computing technology from the late 20[th] century to the present. More and more financial institutions and major market trading are gradually changing from traditional on-site trading to electronic remote trading, and even automated trading can be performed in a pre-programmed computer programs (algorithmic trading). According to statistics, the US stock market with the most developed financial development currently has as many as 60 to 70 percent of its trading automated by programmed programs. In this wave of application of computer computing technology and financial markets, the combination of quantitative trading and computer algorithm is particularly prosperous [1].

The main function of the financial market is to provide the current market price of traded commodities so that market participants can benefit from it. Quantitative trading emerged in the stock market in the late 20$^{th}$ century. In recent years, it has been widely applied in automated trading systems in the stock, currency, and futures markets. In order to receive excess returns, statistics and mathematical models are used to obtain the high probability of the market happening in the future and formulate a corresponding set of modeled trading logic for market trading by observing a large amount of historical data in the past.

The biggest feature of quantitative trading is the use of a fixed set of logic for trading, which is most often used in stock and futures trading market, in order to obtain a stable and continuous excess return above the average return. In addition, the aspects considered by quantitative trading are becoming more and more diverse including market composition, developmental evaluation of investment targets, market sentiment, etc., which can be used for further analysis and reference. With the development of the information age, more and more relevant information has been completely collected and retained, which can provide quantitative trading models for reference and analysis, make up for the inability of the human brain to quickly digest huge amounts of data and can make decisions more objectively and not be affected by market sentiment.

Nowadays, with the changes of the times, international trade and the internet have promoted the active free trade market, which reduces trading costs and expands the scale and scope of trading, and the relationship between the financial industry and technology is also deepening as well as becoming more mature and popular [2, 3]. How to use the increasingly developed computer computing as an auxiliary tool for financial market trading and even becoming the key to the success of traders' profits has become the focus of research in the financial community in recent years.

In the early e-commerce applications, more attention was paid to how to transform the physical trading form into an electronic trading form including inputting and storing trading information by the form of electronic files, which is convenient for future search and analysis. Such simple trading is changed from manual services to electronic services, which greatly reduces labor costs.

After the preliminary results of electronic finance in recent years, financial industry players hope that in addition to using computers to complete simple trading procedures, they can further use mature computer programs to help people complete more complex trading decisions. Especially in recent years, artificial intelligence has become increasingly prosperous, and there have been many breakthrough developments [4-7]. How to use computer programs to imitate human thinking logic, assist trading, and even think faster, wider and deeper than humans has become the direction of development and research in the financial field in recent years.

Trading-related programming algorithms have emerged since the late 1980s. In the early days, more human or statistical tools were used for analysis and decision-making, and then computer programs were used for trading. But today, in the past ten to twenty years, more and more quantitative trading has modeled the decision-making process and designed it into the trading algorithm with the trading logic resulting to make the algorithm more intelligent and automatic. And it has become a highly practical and popular research, which has attracted the attention and practice of many governments and financial institutions [8].

Today, there are many rich and diverse research results in the field of trading algorithm research, such as the use of mean regression to adjust the allocation of stock investment weights [9], the use of long short-term memory (LSTM) combined with grid trading method (GTM) to predict market trends for currency trading [10], using box theory combined with support vector machine (SVM) to assist stock trading decisions [11], using Ichimoku Kinkohyo to analyze foreign exchange trading [12], and the use of market trends instead of fixed time series to make trading decision for the trading strategy of the execution unit [13]. It is not difficult to see that the use of computer programs to assist trading and even directly making trading decisions and executing trading have become an unstoppable trend in financial development.

The most basic definition of grid trading method is an average pending order within a specific price range to carry out a trading strategy of buying low and selling high [14]. The principle of grid trading is simple but it can effectively earn spread profits when prices fluctuate. In recent years, this strategy has been often applied to arbitrage in many financial markets [15-16].

Simplified Swarm Optimization (SSO) was proposed by Yeh [17] in 2009 to improve the problem of Particle Swarm Optimization (PSO) in solving discrete problems with the core concept of simplicity. SSO simplifies the algorithm and improves the efficiency of the solution, and has been widely used in solving problems in many different fields [18-29]. Finding the best parameter combination is one of the SSO applications. Multiple parameters can be considered and adjusted at the same time to find the best parameter combination, which has been used to adjust the hyperparameter combination of convolutional neural network [30], optimize the parameters of solar models [31-33], and adjust the parameters in artificial neural network (ANN) [34].

Recently, the application of deep learning (DL) and ANN in financial activities has become more and more diverse, which the most widely used is the prediction of the stock market, exchange rate, and financial index, etc. [35]. At present, there are still many studies trying to use various neural network models to assist people in financial forecasting and decision-making [36-38] for making more rational and accurate judgments.

To sum up, this study proposed a new grid trading model, adopts SSO algorithm to find parameters suitable for various market situations as input values and labels, trains the fully connected neural network and LSTM model, and finally, a quantitative trading model can automatically calculate and adjust the optimal trading parameters for trading after inputting the existing market situation.

Following the rise of e-commerce and quantitative trading mentioned above, the relationship between people's economic activities and technology has become more and more inseparable, and many trading relies on programs to complete. In future trends, programs will not only help people record trading and complete transfers but even help people make trading decisions automatically including determining the timing and price of trading. Against the foregoing background, the purpose of this study is to:
1. Provide a new set of grid trading algorithms to improve the shortcomings of premature entry and exit

of existing grid trading models in the market.
2. This trading algorithm can adapt to change in the external environment as time and market conditions change, and self-adjust the model to reduce investors' effort in the trading market.
3. Through a set of trading models with logical rules, the irrational decisions brought about by investors' subjective trading decisions are reduced.
4. Balance the relationship between risk and profit, and get a great reward under a certain reasonable risk.

## 2. Overview of grid trading, SSO, and DL

2.1. Grid trading

The most basic definition of grid trading method is an average pending order within a specific price range to carry out a trading strategy of buying low and selling high [14]. There are two key factors in grid trading that determine the effectiveness of this strategy. The first is price volatility including both ups and downs. If the price goes up all the way, those who hold the spot make more profit than grid trading. If the price goes down all the way, spot traders basically cannot have many profit opportunities but short futures buyers are more able to make profits. If the market goes up and down, the grid trading strategy has the greatest profitability. The more frequent the fluctuations, the greater the profit rate increases. In most markets, prices are highly fluctuating and often have a price mean reversion [39-41] so that grid trading strategies are now being used by more and more market players.

The other key determines the profit size of grid trading is its parameter setting including grid average, upper bound of the grid, lower bound of the grid, number of grids, initial price, stop loss point, stop profit point, etc. The setting of the above parameters should be based on product price fluctuations, trading costs, risk tolerance, and the amount of principal as the setting considerations, which directly affect the final performance of grid trading.

Grid trading is mainly divided into two types: equal-distance grid and equal-ratio grid. At present, the grid that many people use in practice is equal-distance grid.

The setting of equal-distance grid is based on the initial price $P_0$, and the setting of upper bound of the grid $G_{ul}$, the lower bound of the grid $G_{ll}$, the total number of grids $n$, and the calculation of the grid spacing $G_s$ by Eq. (1).

$$G_s = \frac{G_{ul} - G_{ll}}{n} \qquad (1)$$

After the setting of the grid model, the trading is based on this grid to buy and sell financial products with the rise and fall of prices. For an example of the equal-distance grid, please refer to Figure 1.

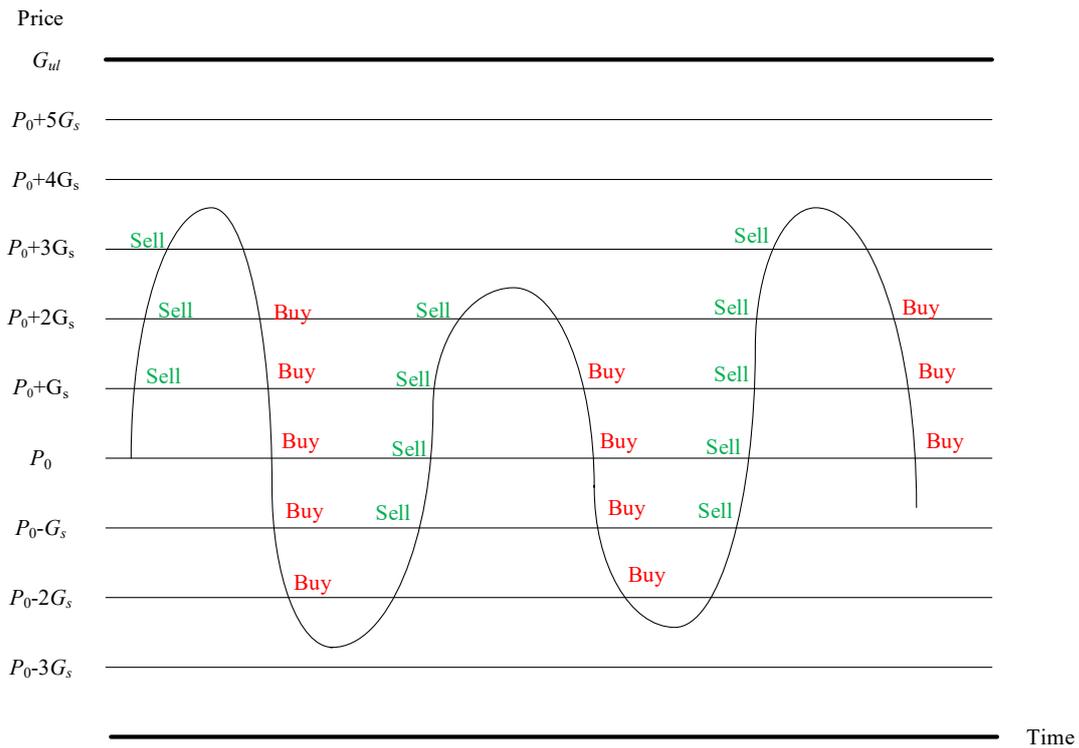

Figure 1. Schematic diagram of equal-distance grid trading

The grid spacing $G_s$ of equal-ratio grid is a fixed ratio and the equal-ratio grid is calculated at this ratio. For an example of the equal- ratio grid, please refer to Figure 2.

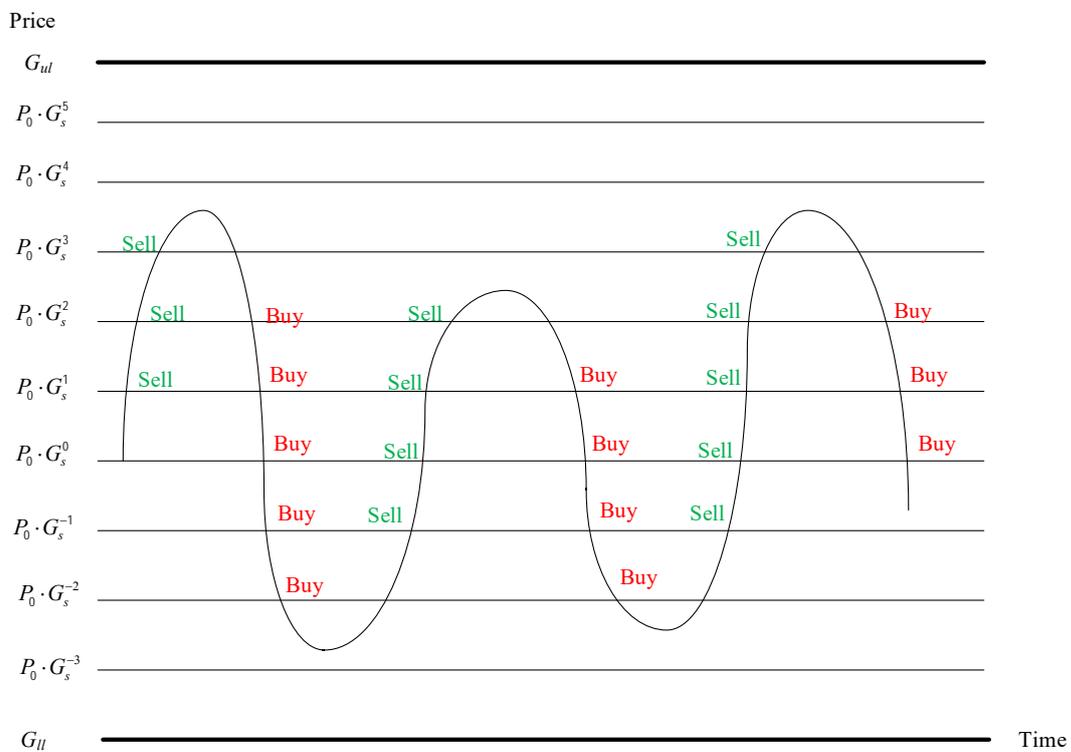

Figure 2. Schematic diagram of equal-ratio grid trading

## 2.2. SSO

Simplified Swarm Optimization (SSO) was proposed by Yeh [17] in 2009 bases on the concepts of personal best (*pbest*) and global best (*gbest*) and adds random numbers resulting that the solution owns the opportunity to escape from the local optimal solution and improve the diversity of the solution.

The major difference between SSO and other heuristic algorithms is its special update mechanism. There are three particularly important parameter settings including $C_g$, $C_p$, and $C_w$, where $C_g > C_p > C_w$. The update mechanism is shown in Eq. (2). According to the correspondence between the random number $\rho$ and $C_g$, $C_p$, and $C_w$, the next generation solution $x_{i,j}^t$ is determined, which may be *gbest*, *pbest*, the current solution or a random number, respectively.

$$x_{i,j}^{t+1} = \begin{cases} g_j, & \text{if } \rho \in [0, C_g) \\ p_{i,j}, & \text{if } \rho \in [C_g, C_p) \\ x_{i,j}^t, & \text{if } \rho \in [C_p, C_w) \\ x, & \text{if } \rho \in [C_w, 1) \end{cases} \quad (2)$$

where, let $x_{i,j}^t = x_{i,1}^t, x_{i,2}^t, \ldots, x_{i,j}^t$ be the *i*th solution with *j* variables in the *t*th generation; $\rho$ is a random number subjects to uniform distribution between [0, 1].

The update method is that when $\rho$ is between [0, $C_g$), the variable $x_{i,j}^{t+1}$ is replaced by the global best solution $g_j$ that is the best performing solution among all the solutions at present; when $\rho$ is between [$C_g$, $C_p$), it is replaced by the best solution in the region $p_{i,j}$ that is the optimal solution in the past generations of the variable; when is between [$C_p$, $C_w$), it maintains the solution of the previous generation $x_{i,j}^t$; when is between [$C_w$, 1), it is replaced by $x$ that is a random number generated in the upper and lower bounds of the variable. The purpose is to reduce the chance of getting trapped in local optimal solutions, while also increasing the diversity of solutions.

## 2.3. DL

The ANN, Back-propagating method, and Long Shout-Term Memory (LSTM) adopted in this study have an overview in this section.

### 2.3.1. ANN

In recent years, DL has been widely used in medicine, industry, transportation and other fields, which can help the completion of speech recognition and machine vision. It is a machine learning (ML) algorithm based on ANN, which learns through data features.

The computing architecture of ANN was first proposed by WS. McCulloch and W. Pitts [42], and subsequently it has been continuously improved by many outstanding scholars [43, 44], and it has become a famous ML model in artificial intelligence (AI).

The basic structure of ANN is divided into three parts including input layer, hidden layer and output layer. Each node of the input layer corresponds to the input predictor variable, each node of the output layer corresponds to the objective variable, and the hidden layer is sandwiched between input layer and output layer.

Except for the nodes in the input layer, each node in the ANN is connected to several nodes in front of it, which is this model is called fully connected neural network (FNN), and each node has a corresponding weight.

2.3.2. Back-propagating method

In DL, the back-propagating method [45] is an extremely important key to make the model complete and is often used to train and optimize ANN. Using the back-propagation, the gradient of the loss function to the weight can be efficiently found, and then the gradient descent method [46] is used to solve each weight. The "loss" in the loss function refers to the error of the actual value and the predicted value. The main concept of back-propagation is to return the error resulting that the weight can use the error size to perform gradient descent method to obtain and update the more suitable weight, further reduce the error and optimize the weight.

2.3.3. LSTM

LSTM is a model generated to improve the short-term memory of recurrent neural network (RNN). It is mainly composed of four units including memory cell, input gate, output gate and forget gate.

The input gate controls whether it is input into the memory unit this time, the memory unit is responsible for storing the calculated value, the forget gate controls to clear the memory, and the output gate controls whether to output the operation result.

**3. Proposed Approach**

The approach proposed in this study is presented in sequence. The initial setting and operation mechanism of grid trading is introduced in Subsection 3.1. Subsection 3.2 presents the concept and setting method of flexible grid. The use of SSO to obtain the optimal flexible grid parameters in different situations is shown in Subsection 3.3. As the training basis for DL, a neural network is finally trained that can automatically output the suitable grid parameters for trading by inputting recent market information to obtain excess returns from market fluctuations is represented in Subsection 3.4.

3.1. Operation mechanism of grid trading

This section presents in detail how grid trading practices in this study, including initial parameter setting and calculation, and subsequent operation mechanisms and processes.

3.1.1. Initial parameter setting of grid trading

A total of five basic parameters need to be set, namely the total investment capital $F_0$, the initial price

$P_0$, the upper bound of the grid $G_{ul}$, the lower bound of the grid $G_{ll}$, and the total number of grids $n$, before operating a grid trading model. The total investment capital and initial price are set as control variables in this study, i.e., when comparing the results, the total investment capital and initial price used in any method will be set to the same fixed value, and compared on the same standard.

Before the grid trading model runs, the unit price difference $G_s$, which is the grid spacing, needs to be calculated using Eq. (1) in Section 2 if the grid is an equal-distance grid. $G_s$ is a certain value in equal-distance grid. But it is a fixed ratio in equal-ratio grid, for example, the value of the next grid is $P_0 \cdot G_s^1 = 110$ and the next grid is $P_0 \cdot G_s^2 = 121$ when $P_0 = 100$ and $G_s = 1.1$. And this ration can be calculated from Eq. (3) below.

$$G_s = \sqrt[n]{\frac{G_{ul} - G_{ll}}{G_{ll}} + 1} \tag{3}$$

In addition, it is also necessary to calculate how much funds should be used to purchase the spot before grid trading, i.e., the initial purchase of spot $S_0$, so as to sell for profit when the price rises. And how much funds must be kept on hand to buy sopt when the price falls, here $C_0$ means to start holding cash. Hence, the total investment funds are divided into two parts that is represented by the following Eq. (4). The spot and funds held in each subsequent period are represented by $S_j$ and $C_j$, where $j$ represents the $j$th period.

$$F_0 = S_0 + C_0 \tag{4}$$

In order to calculate the initial purchase of spot $S_0$ and the initial holding of cash $C_0$, it is necessary to first calculate the number of upper grids $n_u$ and the number of lower grids $n_l$, the sum of which is equivalent to the total number of grids $n$ such as shown in Eq. (5). The values of $n_u$ and $n_l$ can be obtained by Eqs. (6)-(7), respectively. Here, the equal-distance grid is used as an example.

$$n = n_u + n_l \tag{5}$$

$$n_u = (G_{ul} - P_0) / G_s \tag{6}$$

$$n_l = (P_0 - G_{ll}) / G_s \tag{7}$$

After calculating the number of upper grids $n_u$ and the number of lower grids $n_l$, and then further use Eqs. (8)-(9) to calculate the initial purchase of spot $S_0$ and the initial holding of cash $C_0$. At this time, the single-cell trading volume $G_v$ is still an unknown value.

$$S_0 = G_v \cdot n_u \cdot P_0 \tag{8}$$

$$C_0 = G_v \cdot [(P_0 - G_s) + G_{ll}] / 2 \cdot n_l \tag{9}$$

Finally, substitute Eqs. (8)-(9) into Eq. (4) to get the single-cell trading volume $G_v$, which can be calculated by Eq. (10).

$$G_v = F_0 / \{[(P_0 - G_s) + G_{ll}] / 2 \cdot n_l + n_u \cdot P_0\} \tag{10}$$

Finally, calculate the price of each grid cell $g_i$, $i$ from 1 to $n$ using Eq. (11) (equal-distance grid) or Eq. (12) (equal-ratio grid).

$$g_i = G_{ll} + G_s \cdot (i - 1) \tag{11}$$

$$g_i = G_{ll} \cdot G_s^{(i-1)} \tag{12}$$

3.1.2. Operation mechanism of grid trading

The following Figures 3 (*a*)-(*e*) illustrate in detail how the grid trading model updates and adjusts with the market price after the initial parameters are set, and how to settle at the end.

1. When the grid is initially running, the current price is used as the benchmark, and the above grid price is placed on a sell order, and the following grid price is placed on a buy order as shown in Fig. 3 (*a*).

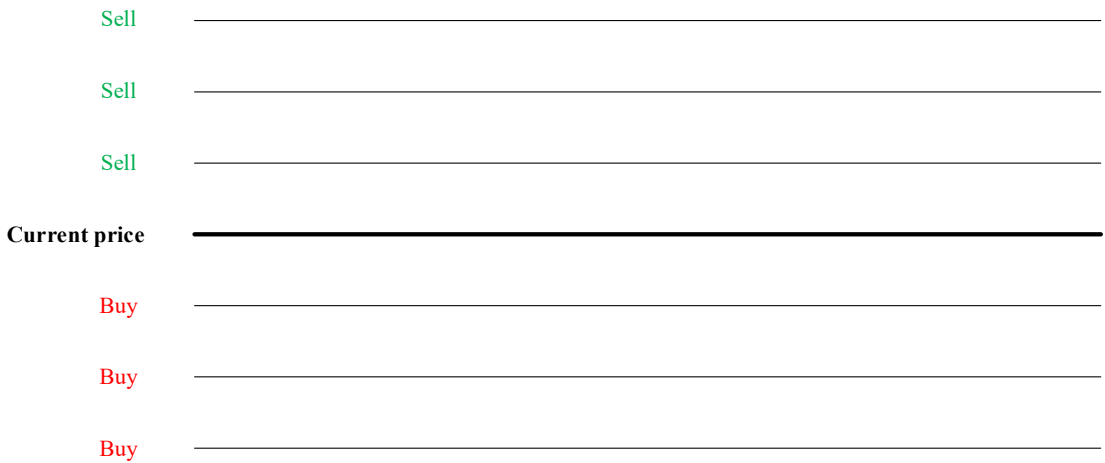

Fig. 3. Grid trading operation chart (*a*)

2. If the price rises until it hits the first grid line, make a sell action, update the spot volume and funds held, and place a buy order at the original grid position as shown in Fig. 3 (*b*).

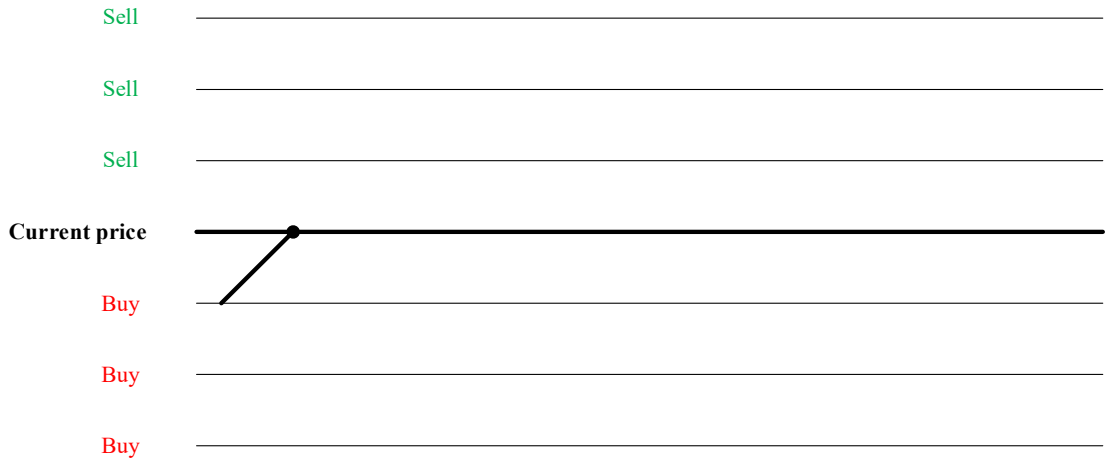

Fig. 3. Grid trading operation chart (*b*)

3. If the price falls back to the initial grid line, make a buy action, update the spot volume and funds held, and place a sell order at the original grid position as shown in Fig. 3 (*c*).

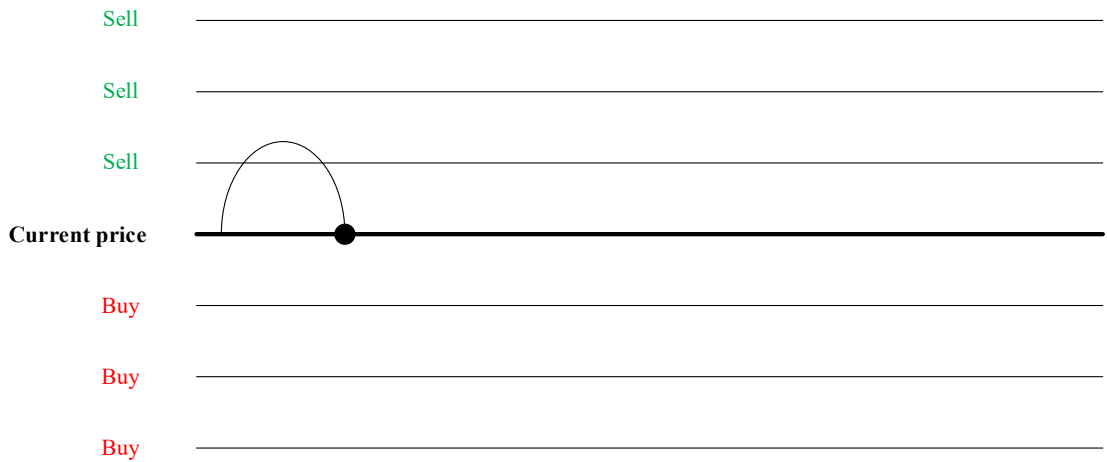

Fig. 3. Grid trading operation chart (*c*)

4. If the price continues to drop to a grid line, make a buy action, update the spot volume and funds held, and place a sell order at the original grid position as shown in Fig. 3 (*d*).

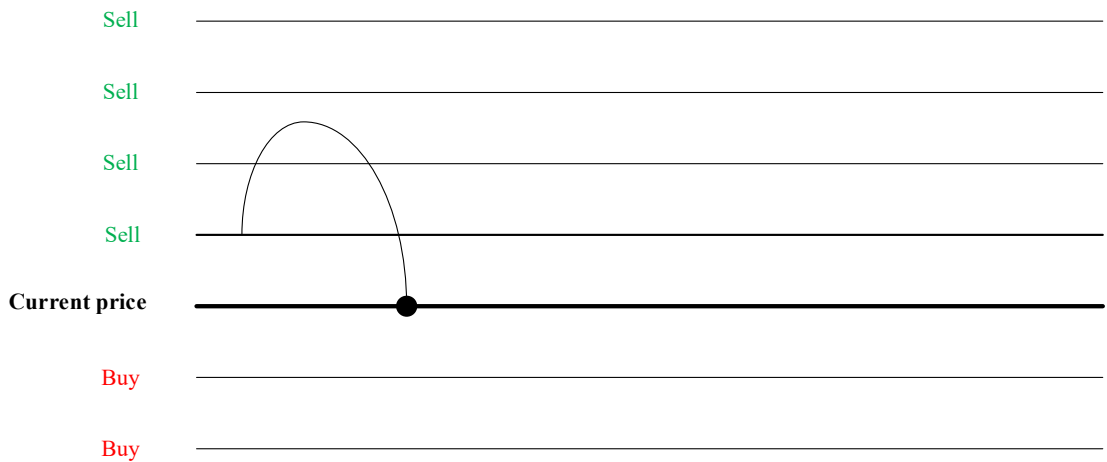

Fig. 3. Grid trading operation chart (*d*)

5. Continue to trade with the above mechanism. Although the price has returned to the original point of grid trading, it has successfully arbitraged 7 times that is equivalent to 7 grids of grid spread profits as shown in Fig. 3 (*e*).

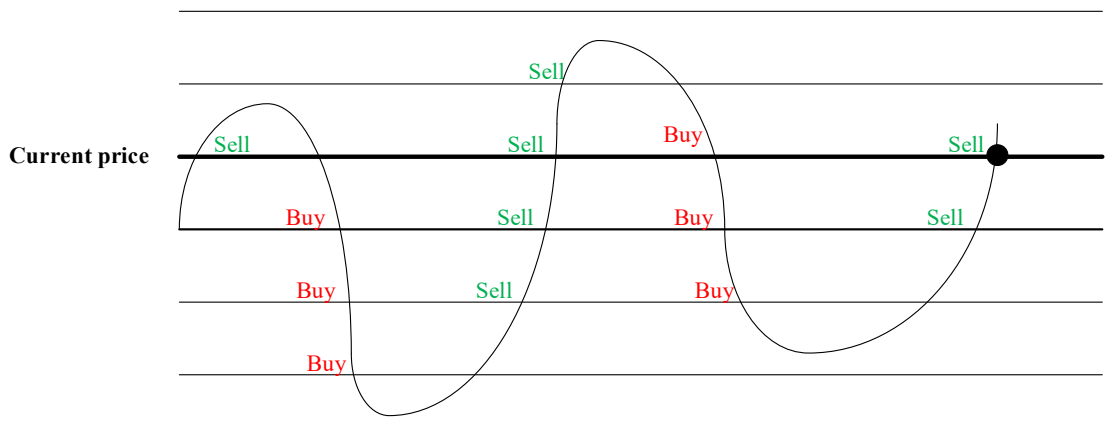

Fig. 3. Grid trading operation chart (*e*)

6. When the grid trading model is to be closed, there are two ways to end it. One is to directly keep the current spot and funds held, and the other is to sell the spot at the current price and convert it into cash. The former is recommended to be used when the market price is low, and the latter is not recommended. In this study, the grid is closed and settled in the second method.

For the overall process of grid trading operation, please refer to the grid trading operation flow chart in following Fig. 4.

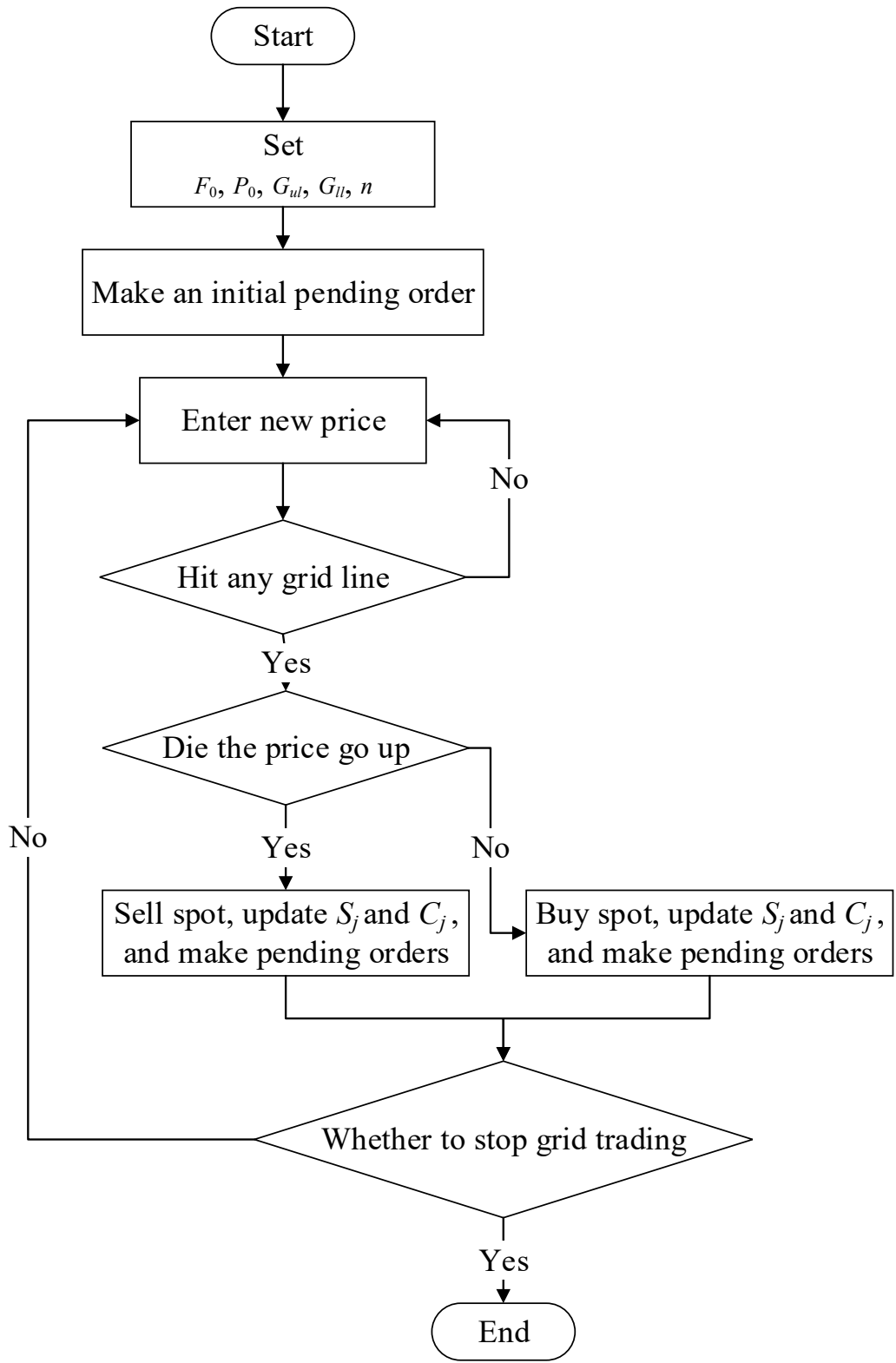

Fig. 4. Grid trading operation flow chart.

3.2. Concept and architecture of flexible grid

Based on the equal-distance and equal-ratio grid trading models used in the financial market today, this study proposes a new and adaptive grid trading model: flexible grid. Through a more flexible grid structure combined with SSO algorithm and DL, a grid trading model that can adjust parameters according to market changes and adapt to various external conditions is constructed.

Under the framework of the equal-distance grid trading model, it can exert its maximum model benefits when the market fluctuates. Even if the price returns to the original point, it is still possible to arbitrage from the volatile market situation based on the adoption of the mean reversion strategy in quantitative trading [47]. On the other hand, equal-ratio grid trading can obtain better returns in the volatile rise based on the combination of the two strategies of mean reversion and rend following [48].

The flexible grid proposed in this study captures the advantages of the equal-distance grid and the equal-ratio grid at the same time. It can outperform the traditional grid trading structure of the past whether the market is sideways, rising or even falling.

When the flexible grid is initially set, it also needs to set its total investment capital $F_0$, the initial price $P_0$, the upper bound of the grid $G_{ul}$, the lower bound of the grid $G_{ll}$, the number of upper grids $n_u$ and the number of lower grids $n_l$. There are two main differences compared with other models in the initial setting:

1. The number of upper grids $n_u$ and the number of lower grids $n_l$ in Eq. (5) are no longer calculated with Eq. (6) and Eq. (7) but can be set initially.
2. The grid is divided into upper and lower parts with the initial price $P_0$ as the boundary. The upper part and the lower part can set the number of the grids and have their own grid spacing ratio. The ratio of the upper grid spacing is $G_{su}$ and the lower grid spacing is $G_{sl}$. It should be noted here that $G_{su}$ must be a number greater than 0 and less than 1, and $G_{sl}$ must be a number greater than 1. The feature of this is that the upper grid spacing becomes smaller and smaller as the price is higher, i.e., the trading frequency becomes more and more frequent. Similarly, the lower grid spacing becomes smaller and denser when the price is lower

For the specific calculation method of grid value, please refer to Eqs. (13)-(14) and the over structure can refer to Figure 5 below.

$$G_{su} = \frac{1}{\sqrt[n_u]{\frac{G_{ul} - P_0}{P_0} + 1}} \quad (13)$$

$$G_{sl} = \sqrt[n_l]{\frac{P_0 - G_{ll}}{G_{ll}} + 1} \quad (14)$$

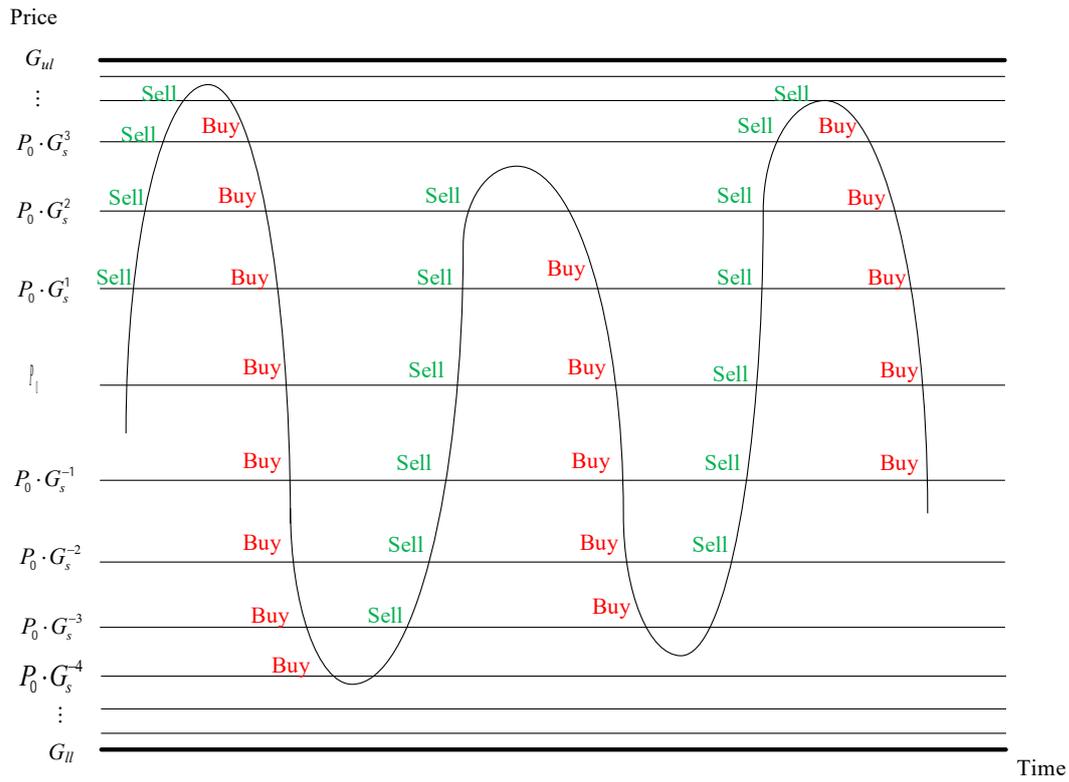

Figure 5. Schematic diagram of flexible grid trading.

The core concept in the design process of flexible grid is to perform more frequent sell actions when the price is high, and buy the cheaper spot more frequently when the price is lower. The flexible grid has a similar structure to the equal-distance grid in the middle section of the model, while the lower half of the model is more similar to the equal-ratio grid, and the upper half of the model is different from the current common equal-distance grid and equal-ratio grid trading model with higher mesh density at higher prices. The main purpose of this model is to capture the advantages of equal-distance grid and equal-ratio grid, and can have better performance than the current existing model whether the market trend is volatile, continuous rise or fall.

However, in order to truly take advantage of the flexible grid model architecture, optimal parameter setting and self-adjustment in line with the market situation are crucial determinants for the success or failure of the model. Therefore, this study uses a SSO algorithm to determine the appropriate parameters of the flexible grid in different situations that is described in Section 3.3, and input the calculated parameter combination and the corresponding market situation into the artificial neural network (ANN) for model training that is presented in Section 3.4. Finally, this study produces a trained DL model, which can automatically adjust to the grid parameters most suitable for the current market situation by simply inputting the current market situation.

3.3. SSO for optimal parameters

In this subsection, the SSO algorithm is adopted to adjust the optimal solution of the flexible grid under different market conditions based on the flexible grid constructed in the previous subsection. The market situation and the optimal parameters are input into the ANN as the training set for model training in subsection 3.4.

3.3.1. Objective and Constraint

The main objective pursued by this study is to maximize the investment return. Therefore, the market investment commodity price is set by the SSO algorithm after the j period of update. And the objective model please refer to following Eq. (15).

$$\text{Max} \quad S_j \times P_j \times C_j \tag{15}$$

where $S_j$, $P_j$, and $C_j$ represent the spot quantity, the commodity market price, and the funds held in the last period after the initial purchase of spot $S_0$ and initial holding of funds $C_0$ are inputted through the price of a total j-periods.

In addition, in the flexible grid model, it is necessary to ensure that the profit of each trading is greater than the trading cost $h\%$ (usually the trading fee rate) so that the following constraint Eq. (16) is set, where $i$ is from 0 to $n$.

$$s.t. \quad g_{i+1} - g_i > h\% \times g_{i+1} \tag{16}$$

The calculation of $g_i$ can refer to Eqs. (11)- (12). After substituting it, it can be found that this constraint is actually a constraint on the grid parameters $G_{ul}$, $G_{ll}$, $n_u$, and $n_l$.

On the other hand, additional constraints are placed on the grid upper and lower bounds $G_{ul}$, $G_{ll}$, $n_u$, and $n_l$ for experiment 1 and experiment 2 as shown in Eqs. (17)- (19), where $x$ is from 0 to $j$.

$$s.t. \text{ (experiment 1)} \quad P_0 \times 105\% < G_{ul} < P_0 \times 130\% \tag{17}$$

$$P_0 \times 70\% < G_{ll} < P_0 \times 95\% \tag{18}$$

$$s.t. \text{ (experiment 2)} \quad P_0 \times 105\% < G_{ul} < P_0 \times 150\% \tag{17}$$

$$P_0 \times 50\% < G_{ll} < P_0 \times 95\% \tag{18}$$

$$10 < n_l < [P_0 \times 100/(maximum\ P_x \times 1.3)] - 10 \tag{19}$$

The above four constraints are to keep the error space and avoid the overfitting of subsequent ANN training, which cause the price to easily exceed the upper and lower limits of the grid resulting to lose arbitrage opportunities, and control the parameters within a reasonable range.

3.3.2. Solution encoding

According to subsection 3.3.1, the construction of a grid trading model must be set by the total investment $F_0$, initial price $P_0$, grid upper bound $G_{ul}$, grid lower bound $G_{ll}$ and total number of grids $n$. In this study, $F_0$ and $P_0$ are control variables, hence, the parameters to be solved by SSO are $G_{ul}$, $G_{ll}$ and $n$. In flexible grid, the total number of grids $n$ can be divided into the number of upper grids $n_u$ and the number of lower grids $n_l$.

And the solution range can refer to Table 1 below.

Table 1. Upper and lower bounds of grid trading parameters

| Scope of use | Parameters | Variable upper bound | Variable lower bound |
|---|---|---|---|
| Experiment 1 | grid upper bound $G_{ul}$ | $P_0 \times 130\%$ | $P_0 \times 105\%$ |
| | grid lower bound $G_{ll}$ | $P_0 \times 95\%$ | $P_0 \times 70\%$ |
| Experiment 2 | grid upper bound $G_{ul}$, | $P_0 \times 150\%$ | $P_0 \times 105\%$ |
| | grid lower bound $G_{ll}$ | $P_0 \times 95\%$ | $P_0 \times 50\%$ |
| In common use | number of upper grids $n_u$ | $[P_0 \times 100/(maximum\ P_x \times 1.3)] - 10$ | 10 |
| | number of lower grids $n_l$ | $[P_0 \times 100/(maximum\ P_x \times 1.3)] - 10$ | 10 |

The solution encoding in this study please refer to following Figure 6, which $x_1$, $x_2$, $x_3$, and $x_4$ are set as the grid upper bound $G_{ul}$, the grid lower bound $G_{ll}$, the number of upper grids $n_u$, and the number of lower grids $n_l$, respectively.

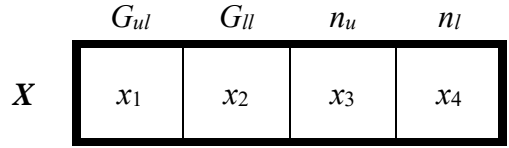

Figure 6. Structure of solution encoding

For example, when $X = (150, 100, 10, 30)$, it means the grid upper bound is 150, the grid lower bound is 100, the number of upper grids is 10, the number of lower grids is 30, and the total number of grids is 40.

3.3.3. Parameter setting and scope of update mechanism

This study adopts SSO algorithm to solve the studied problem, where $\rho$ is a random number subjects to uniform distribution between [0, 1], and the solution of the problem is updated according to this random number: when $\rho$ is between [0, $C_g$), the variable $x_{i,j}^{t+1}$ is replaced by the global best solution $g_j$ that is the best performing solution among all the solutions at present; when $\rho$ is between [$C_g$, $C_p$), it is replaced by the best solution in the region $p_{i,j}$ that is the optimal solution in the past generations of the variable; when is between [$C_p$, $C_w$), it maintains the solution of the previous generation $x_{i,j}^t$; when is between [$C_w$, 1), it is replaced by $x$ that is a random number generated in the upper and lower bounds of the variable. The update mechanism can refer to Eq. (2).

The symbols and definitions used in the SSO operation are shown as following Table 2.

Table 2. Symbols and definitions of SSO

| Symbols | Definitions |
| --- | --- |
| $N_{var}$ | The number of variables: the grid upper bound $G_{ul}$, grid lower bound $G_{ll}$ and total number of grids $n$ in this study. |
| $N_{sol}$ | Total number of solutions. |
| $x_i^t$ | $x_i^t = (G_{uli}^t, G_{lli}^t, n_i^t)$ represents the $i$th solution in the $t$th generation, where $t=1, 2,…, N_{gen}$, $i=1, 2, …, N_{sol}$. |
| $\widehat{G_{ul}^p}, \widehat{G_{ul}^g}, \widehat{G_{ll}^p}, \widehat{G_{ll}^g}, \widehat{n^p}, \widehat{n^g}$ | $pbest$ and $gbest$ of each variable during the update process. |
| $C_g, C_p, C_w$ | The three key parameters used to determine the update value in SSO Can be adjusted according to different situations. |
| UB | $UB = (ub_1, ub_2, …, ub_{Nvar})$ is the upper bound of each variable, i.e., $x \leq ub_j$. |
| LB | $LB = (lb_1, lb_2, …, lb_{Nvar})$ is the lower bound of each variable, i.e., $x \geq lb_j$. |

3.4. Training ANN to automatically adjust flexible grid parameters

After using SSO to obtain the optimal grid configuration under various market conditions, the market conditions and the calculated grid parameters are used as the questions and answers for training ANN training. The market situation is interpreted in terms of the following values and input into the ANN, namely, the highest price during the period, the lowest price during the period, the average market price, the average trading quantity, the price change (initial price minus final price), the trading quantity change (initial quantity minus final quantity), price standard deviation, and trading quantity standard deviation.

The output of the ANN outputs the relevant parameters required to construct a grid trading model including upper bound of the grid, lower bound of the grid, the number of upper grids, and the number of lower grids. The specific ANN architecture can refer to the following Figure 7.

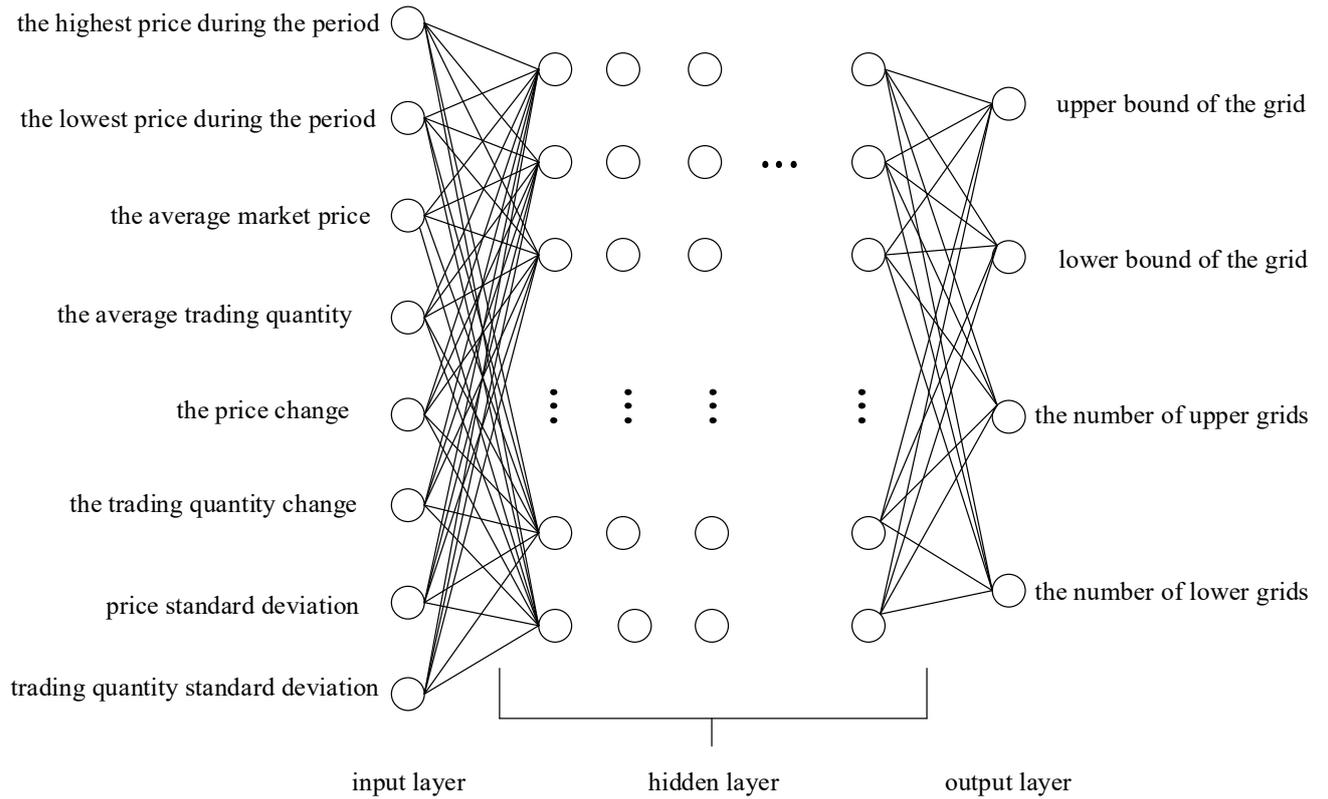

Figure 7. ANN architecture

In the neural network training process, the weights in the ANN nodes continuously adjust themselves through the error value calculated by the loss function until the training is completed when it converges to a state with the smallest error. Afterwards, through this trained neural network, the optimal parameters of the grid trading model can be generated by inputting the recent market state and trend, and a flexible grid model can be automatically constructed for market trading activities.

In this study, two kinds of neural networks are used for training, namely, Fully-connect Neural Network (FNN) and Long Shout-Term Memory (LSTM) neural network, which is often used for time series prediction in recent years, to observe and compare which neural networks perform better in learning grid parameters.

**4. Experimental results**

The data sets used for validation and comparison in this study are Standard & Poor's 500 (S&P 500), NASDAQ Composite, Dow Jones Industrial Average (DJIA), Euro Stoxx 50, and Shanghai Composite, a total of five large-cap indices from 2011 to 2022.

4.1. Verification of flexible grid performance with fixed parameters

First of all, the performance of flexible grid is compared with equal-distance grid and equal-ratio grid based on the same grid number and upper and lower bounds. The results can be found in Tables 3-5. The number of grids is calculated by Eq. (20) to be close to the real investment situation and to ensure the adequacy of the use of funds. According to the upper and lower bound conditions, each data set is divided into two groups for experiments. The first group uses 1.3 times the initial price as the upper bound of the grid, and 0.7 times the initial price as the lower bound of grid; while the second group uses 1.5 times the initial price as the upper bound of the grid, and 0.5 times the initial price as the lower bound of grid.

From the results in Tables 3-5, it can be verified that the flexible grid obviously obtains the best ROI, accumulated wealth and Sharpe ratio through trading with the same grid parameters under ten-year fluctuations of five different composite indices. Overall, flexible grid appropriately delays the entry and exit timing due to its trading structure, and obtains a relatively high ROI and Sharpe ratio.

Table 3. ROI obtained by flexible grid, equal-distance grid, and equal-ratio grid

| Grid type | Flexible grid | Equal-distance | Equal-ratio |
|---|---|---|---|
| **S&P 500** | | | |
| $(G_{ul}, G_{ll}) = (P_0 \times 1.3, P_0 \times 0.7)$ | **31.692 %** | 27.934 % | 22.234 % |
| $(G_{ul}, G_{ll}) = (P_0 \times 1.5, P_0 \times 0.5)$ | **43.727 %** | 39.556 % | 30.174 % |
| **Nasdaq 100** | | | |
| $(G_{ul}, G_{ll}) = (P_0 \times 1.3, P_0 \times 0.7)$ | **53.795 %** | 49.261 % | 40.369 % |
| $(G_{ul}, G_{ll}) = (P_0 \times 1.5, P_0 \times 0.5)$ | **71.888 %** | 66.024 % | 51.180 % |
| **Dow Jones Industrial Average** | | | |
| $(G_{ul}, G_{ll}) = (P_0 \times 1.3, P_0 \times 0.7)$ | **21.935 %** | 18.332 % | 13.629 % |
| $(G_{ul}, G_{ll}) = (P_0 \times 1.5, P_0 \times 0.5)$ | **33.541 %** | 29.741 % | 21.648 % |
| **Euro Stoxx 50** | | | |
| $(G_{ul}, G_{ll}) = (P_0 \times 1.3, P_0 \times 0.7)$ | **0.601 %** | -4.122 % | -7.284 % |
| $(G_{ul}, G_{ll}) = (P_0 \times 1.5, P_0 \times 0.5)$ | **17.499 %** | 12.520 % | 7.087 % |
| **Shanghai Composite** | | | |
| $(G_{ul}, G_{ll}) = (P_0 \times 1.3, P_0 \times 0.7)$ | **-33.904 %** | -38.944 % | -38.534 % |
| $(G_{ul}, G_{ll}) = (P_0 \times 1.5, P_0 \times 0.5)$ | **-12.955 %** | -19.936 % | -18.290 % |

Table 4. Accumulated wealth obtained by flexible grid, equal-distance grid, and equal-ratio grid

| Grid type | Flexible grid | Equal-distance | Equal-ratio |
|---|---|---|---|
| **S&P 500** | | | |
| $(G_{ul}, G_{ll}) = (P_0 \times 1.3, P_0 \times 0.7)$ | **13169** | 12793 | 12223 |
| $(G_{ul}, G_{ll}) = (P_0 \times 1.5, P_0 \times 0.5)$ | **14373** | 13956 | 13017 |
| **Nasdaq 100** | | | |
| $(G_{ul}, G_{ll}) = (P_0 \times 1.3, P_0 \times 0.7)$ | **15380** | 14926 | 14037 |
| $(G_{ul}, G_{ll}) = (P_0 \times 1.5, P_0 \times 0.5)$ | **17189** | 16602 | 15118 |
| **Dow Jones Industrial Average** | | | |
| $(G_{ul}, G_{ll}) = (P_0 \times 1.3, P_0 \times 0.7)$ | **12194** | 11833 | 11363 |
| $(G_{ul}, G_{ll}) = (P_0 \times 1.5, P_0 \times 0.5)$ | **13354** | 12974 | 12165 |
| **Euro Stoxx 50** | | | |
| ($G_{ul}$,$G_{ll}$) = (P₀×1.3,P₀×0.7) | **10060** | 9588 | 9272 |
| ($G_{ul}$,$G_{ll}$) = (P₀×1.5,P₀×0.5) | **11750** | 11252 | 10709 |
| **Shanghai Composite** | | | |
| ($G_{ul}$,$G_{ll}$) = (P₀×1.3,P₀×0.7) | **6610** | 6106 | 6147 |
| ($G_{ul}$,$G_{ll}$) = (P₀×1.5,P₀×0.5) | **8705** | 8006 | 8171 |

Table 5. Sharpe ratio obtained by flexible grid, equal-distance grid, and equal-ratio grid

| Grid type | Flexible grid | Equal-distance | Equal-ratio |
|---|---|---|---|
| **S&P 500** | | | |
| ($G_{ul}$,$G_{ll}$) = (P₀×1.3,P₀×0.7) | **0.118** | 0.103 | 0.092 |
| ($G_{ul}$,$G_{ll}$) = (P₀×1.5,P₀×0.5) | **0.160** | 0.145 | 0.137 |
| **Nasdaq 100** | | | |
| ($G_{ul}$,$G_{ll}$) = (P₀×1.3,P₀×0.7) | **0.187** | 0.172 | 0.161 |
| ($G_{ul}$,$G_{ll}$) = (P₀×1.5,P₀×0.5) | **0.234** | 0.219 | 0.215 |
| **Dow Jones Industrial Average** | | | |
| ($G_{ul}$,$G_{ll}$) = (P₀×1.3,P₀×0.7) | **0.079** | 0.065 | 0.054 |
| ($G_{ul}$,$G_{ll}$) = (P₀×1.5,P₀×0.5) | **0.119** | 0.105 | 0.095 |
| **Euro Stoxx 50** | | | |
| ($G_{ul}$,$G_{ll}$) = (P₀×1.3,P₀×0.7) | **0.002** | -0.011 | -0.021 |
| ($G_{ul}$,$G_{ll}$) = (P₀×1.5,P₀×0.5) | **0.046** | 0.033 | 0.023 |
| **Shanghai Composite** | | | |
| ($G_{ul}$,$G_{ll}$) = (P₀×1.3,P₀×0.7) | **-0.082** | -0.095 | -0.106 |
| ($G_{ul}$,$G_{ll}$) = (P₀×1.5,P₀×0.5) | **-0.030** | -0.047 | -0.053 |

4.2. SSO parameter setting

Part of the dataset is then used to select appropriate SSO parameters for all subsequent experiments.

In the first set of experiments, $C_g$, $C_p$, $C_w$ are configured as shown in Table 6. The meaning is to divide the range of random number $\rho$ into four parts according to 7:1:1:1, and the part with ration 7 is assigned to *gbest*, *pbest*, $x_{i,j}^{t+1}$ and a new solution in turn in the experiment to detect what kind of the solution has more critical influence on producing better quality solutions.

Table 6. Best and worst solutions under different parameter combinations (Experiment 1)

| ($C_g$, $C_p$, $C_w$) | Maximum scale solution | **ROI (max)** | **ROI (min)** |
|---|---|---|---|
| (0.7,0.8,0.9) | gbest | **69.74%** | 66.83% |
| (0.1,0.8,0.9) | *pbest* | 69.62% | **67.18%** |

| ($C_g$, $C_p$, $C_w$) | Maximum scale solution | ROI (max) | ROI (min) |
|---|---|---|---|
| (0.1,0.2,0.9) | $x_{i,j}^{t+1}$ | 69.69% | 66.01% |
| (0.1,0.2,0.3) | new solution | 69.41% | 66.69% |

It can be found from Table 6 that it can produce a better quality solution when *gbest* is the largest, and it is robust at the same time. Therefore, the probability of *gbest* is set to the maximum in the final parameter configuration.

After it is determined that *gbest* is set as the maximum probability, the range of random number $\rho$ is then divided into four parts according to 5:3:1:1. Among them, the probability of taking *gbest* as the solution is set to the maximum 0.5 because *gbest* has been determined in the previous step as the key factor to generate a good quality solution, and the probability of 0.3 is assigned to *pbest*, $x_{i,j}^{t+1}$ and a new solution in turn. According to the experimental results, which solution is the second key factor to produce a better solution, the results are shown in Table 7.

It can be found from Table 7 that it can produce a better quality solution when the new random solution is larger, and it is robust at the same time. Therefore, the probability of the new random solution is set to the next largest in the final parameter configuration.

Table 7. Best and worst solutions under different parameter combinations (Experiment 2)

| ($C_g$, $C_p$, $C_w$) | Maximum scale solution | ROI (max) | ROI (min) |
|---|---|---|---|
| (0.5,0.8,0.9) | *pbest* | 69.77% | 66.22% |
| (0.5,0.6,0.9) | $x_{i,j}^{t+1}$ | 69.77% | 66.91% |
| (0.5,0.6,0.7) | new solution | **69.86%** | **67.83%** |

After it is determined that the new random solution is set as the next largest probability, the range of random number $\rho$ is then divided into four parts according to 3:3:3:1. The subsequent steps are analogous to the first two steps, and Table 8 is obtained as follows. It can be found that $x_{i,j}^{t+1}$ and *pbest* have little difference in the quality of the solution, thus, both probabilities are set to the minimum.

Table 8. Best and worst solutions under different parameter combinations (Experiment 3)

| ($C_g$, $C_p$, $C_w$) | Maximum scale solution | ROI (max) | ROI (min) |
|---|---|---|---|
| (0.3,0.6,0.7) | *pbest* | 69.85% | **67.99%** |
| (0.3,0.4,0.7) | $x_{i,j}^{t+1}$ | **69.90%** | 67.81% |

### 4.3. Verification of flexible grid performance with parameters selected by SSO

After setting the SSO parameters, the flexible grid, the equal-distance grid and the equal-ratio grid are connected to the parameters selected by SSO, respectively. And find solutions with 10 runs, 20 generations per run, and 100 sets of solutions per generation. Two sets of experiments, which is same as the previous verification of flexible grid architecture, are also conducted with the upper and lower bounds of different risk levels. The results are shown in Tables 9-11.

Table 9. ROI obtained by flexible grid, equal-distance grid, and equal-ratio grid with SSO parameters

| Grid type | Flexible grid | Equal-distance | Equal-ratio |
|---|---|---|---|
| **S&P 500** | | | |
| $(G_{ul}, G_{ll}) = (P_0 \times 1.3, P_0 \times 0.7)$ | **82.859 %** | 66.384 % | 58.394 % |
| $(G_{ul}, G_{ll}) = (P_0 \times 1.5, P_0 \times 0.5)$ | **90.269 %** | 77.198 % | 60.774 % |
| **Nasdaq 100** | | | |
| $(G_{ul}, G_{ll}) = (P_0 \times 1.3, P_0 \times 0.7)$ | **111.649 %** | 66.384 % | 82.662 % |
| $(G_{ul}, G_{ll}) = (P_0 \times 1.5, P_0 \times 0.5)$ | **127.268 %** | 110.179 % | 86.208 % |
| **Dow Jones Industrial Average** | | | |
| $(G_{ul}, G_{ll}) = (P_0 \times 1.3, P_0 \times 0.7)$ | **74.509 %** | 60.591 % | 51.893 % |
| $(G_{ul}, G_{ll}) = (P_0 \times 1.5, P_0 \times 0.5)$ | **80.559 %** | 70.063 % | 54.760 % |
| **Euro Stoxx 50** | | | |
| $(G_{ul}, G_{ll}) = (P_0 \times 1.3, P_0 \times 0.7)$ | **77.220 %** | 56.650 % | 48.365 % |
| $(G_{ul}, G_{ll}) = (P_0 \times 1.5, P_0 \times 0.5)$ | **88.844 %** | 72.293 % | 55.856 % |
| **Shanghai Composite** | | | |
| $(G_{ul}, G_{ll}) = (P_0 \times 1.3, P_0 \times 0.7)$ | **58.389 %** | 39.178 % | 33.363 % |
| $(G_{ul}, G_{ll}) = (P_0 \times 1.5, P_0 \times 0.5)$ | **80.954 %** | 58.934 % | 43.639 % |

Table 10. Accumulated wealth obtained by flexible grid, equal-distance grid, and equal-ratio grid with SSO parameters

| Grid type | Flexible grid | Equal-distance | Equal-ratio |
|---|---|---|---|
| **S&P 500** | | | |
| $(G_{ul}, G_{ll}) = (P_0 \times 1.3, P_0 \times 0.7)$ | **18286** | 16638 | 15839 |
| $(G_{ul}, G_{ll}) = (P_0 \times 1.5, P_0 \times 0.5)$ | **19027** | 17720 | 16077 |
| **Nasdaq 100** | | | |
| $(G_{ul}, G_{ll}) = (P_0 \times 1.3, P_0 \times 0.7)$ | **21165** | 16638 | 18266 |
| $(G_{ul}, G_{ll}) = (P_0 \times 1.5, P_0 \times 0.5)$ | **22727** | 21018 | 18621 |
| **Dow Jones Industrial Average** | | | |
| $(G_{ul}, G_{ll}) = (P_0 \times 1.3, P_0 \times 0.7)$ | **17451** | 16059 | 15189 |
| $(G_{ul}, G_{ll}) = (P_0 \times 1.5, P_0 \times 0.5)$ | **18056** | 17006 | 15476 |
| **Euro Stoxx 50** | | | |
| ($G_{ul}$,$G_{ll}$) = (P$_0$×1.3,P$_0$×0.7) | **17722** | 15665 | 14837 |
| ($G_{ul}$,$G_{ll}$) = (P$_0$×1.5,P$_0$×0.5) | **18884** | 17229 | 15586 |
| **Shanghai Composite** | | | |
| ($G_{ul}$,$G_{ll}$) = (P$_0$×1.3,P$_0$×0.7) | **15839** | 13918 | 13336 |
| ($G_{ul}$,$G_{ll}$) = (P$_0$×1.5,P$_0$×0.5) | **18095** | 15893 | 14364 |

Table 11. Sharpe ratio obtained by flexible grid, equal-distance grid, and equal-ratio grid with SSO parameters

| Grid type | Flexible grid | Equal-distance | Equal-ratio |
|---|---|---|---|
| **S&P 500** | | | |
| ($G_{ul}$,$G_{ll}$) = (P$_0$×1.3,P$_0$×0.7) | **0.439** | 0.350 | 0.351 |
| ($G_{ul}$,$G_{ll}$) = (P$_0$×1.5,P$_0$×0.5) | **0.442** | 0.391 | 0.388 |
| **Nasdaq 100** | | | |
| ($G_{ul}$,$G_{ll}$) = (P$_0$×1.3,P$_0$×0.7) | **0.498** | 0.350 | 0.437 |
| ($G_{ul}$,$G_{ll}$) = (P$_0$×1.5,P$_0$×0.5) | **0.524** | 0.464 | 0.462 |
| **Dow Jones Industrial Average** | | | |
| ($G_{ul}$,$G_{ll}$) = (P$_0$×1.3,P$_0$×0.7) | **0.377** | 0.310 | 0.301 |
| ($G_{ul}$,$G_{ll}$) = (P$_0$×1.5,P$_0$×0.5) | **0.412** | 0.344 | 0.341 |
| **Euro Stoxx 50** | | | |
| ($G_{ul}$,$G_{ll}$) = (P$_0$×1.3,P$_0$×0.7) | **0.300** | 0.219 | 0.211 |
| ($G_{ul}$,$G_{ll}$) = (P$_0$×1.5,P$_0$×0.5) | **0.331** | 0.265 | 0.257 |
| **Shanghai Composite** | | | |
| ($G_{ul}$,$G_{ll}$) = (P$_0$×1.3,P$_0$×0.7) | **0.189** | 0.134 | 0.131 |
| ($G_{ul}$,$G_{ll}$) = (P$_0$×1.5,P$_0$×0.5) | **0.220** | 0.179 | 0.172 |

From the results in Tables 9-11, it can be verified that in the application of SSO to solve the grid trading parameters, the flexible grid is still the best model in terms of ROI, accumulated wealth and Sharpe ratio compared with the equal-distance grid and equal-ratio grid. In addition, comparing the grid trading results connected with the SSO parameters and the fixed parameter version, it can be found that the ROI has increased significantly. The overall ROI results, the flexible grid is the best, the equal-distance grid is the second, and the equal-ratio grid is the worst.

4.4. Training ANN to automatically adjust flexible grid parameters

After confirming the excellent performance of flexible grid using SSO to search for trading parameters, the flexible grid and SSO are adopted to record the best trading parameters of each index in ten years, and obtain a set of trading parameters every 30 days. In order to expand the follow-up training and data, the moving pace of the model is set to 5. In each index, about 500 pieces of parameter data are obtained for the training and verification of ANN, in which the data of the first nine years is used as the training set and the data of the last year is used as the validation set.

Because the data of the training set in this study has temporal continuity, the data is re-randomly ordered before training in the FNN in order to avoid affecting the training results of the model. In the LSTM neural network, the input training data must be sequential because the model design, thus, the training data does not need to be re-ordered randomly.

The architecture of the FNN and the LSTM neural network has undergone many experiments, and the final architecture and hyper-parameter settings are shown in Tables 12-14.

The input parameters include the highest price during the period, the lowest price during the period, the average market price, the average trading quantity, the price change, the trading quantity change, price standard deviation, and trading quantity standard deviation, which a total of eight variables to describe a market situation. And the output variables are set to upper bound of the grid, lower bound of the grid, the number of upper grids, and the number of lower grids for grid trading.

The number of hidden layers is set to three. The optimizer is set to adam, which is common in recent years, has fast convergence speed, and has excellent performance in searching solution. The excitation function uses sigmoid and relu. And the loss function adopts mean squared error, which is commonly used in regression problems.

Table 12. FNN hyper-parameters and architecture related settings

| Item | Value set |
| --- | --- |
| Number of input variables | 8 |
| Number of hidden layers | 3 |
| Number of output variables | 4 |
| Number of hidden layer nodes | 500 |
| Optimizer | adam |
| Excitation function | sigmoid |
| Loss function | mean squared error |
| Number of generations | 300 |
| Batch size | 40 |

Table 13. LSTM neural network hyper-parameters and architecture related settings

| Item | Value set |
| --- | --- |
| Number of input variables | 8 |
| Number of hidden layers | 3 |
| Number of output variables | 4 |
| Number of hidden layer nodes | 256, 128, 64 |
| Optimizer | adam |
| Excitation function | relu |
| Loss function | mean squared error |
| Number of generations | 300 |
| Batch size | 32 |

After the neural network training set above, the training and comparison results are presented in Tables 14-15. From the results of the mean square error as shown in Eq. (20) in Table 14: Except for the upper and lower bound of the grid in the Nasdaq 100 index, LSTM has a smaller mean square error. All four output variables in the other indices have a smaller mean square error, which the mean square errors of the four output variables in other indices are all FNN performs better.

$$MSE = \frac{1}{N}\sum_{i=1}^{N}(y_i - \hat{y}_i)^2 \qquad (20)$$

Table 14. Comparison of mean square error between FNN and LSTM models

| Neural network type | FNN | LSTM |
| --- | --- | --- |
| **S&P 500** | | |
| upper bound of the grid $G_{ul}$ | **129510** | 186366 |
| lower bound of the grid $G_{ll}$ | **59787** | 108622 |
| number of upper grids $n_u$ | **207** | 252 |
| number of lower grids $n_l$ | **308** | 442 |
| **Nasdaq 100** | | |
| upper bound of the grid $G_{ul}$ | 4725772 | **518181** |
| lower bound of the grid $G_{ll}$ | 1681699 | **1162381** |
| number of upper grids $n_u$ | **133** | 296 |
| number of lower grids $n_l$ | **246** | 568 |
| **Dow Jones Industrial Average** | | |
| upper bound of the grid $G_{ul}$ | **6016533** | 24446812 |
| lower bound of the grid $G_{ll}$ | **2011354** | 19594848 |
| number of upper grids $n_u$ | **164** | 237 |
| number of lower grids $n_l$ | **319** | 402 |
| **Euro Stoxx 50** | | |
| upper bound of the grid $G_{ul}$ | **105024** | 334456 |
| lower bound of the grid $G_{ll}$ | **33988** | 106422 |
| number of upper grids $n_u$ | **132** | 202 |
| number of lower grids $n_l$ | **199** | 290 |
| **Shanghai Composite** | | |
| upper bound of the grid $G_{ul}$ | **65199** | 92869 |
| lower bound of the grid $G_{ll}$ | 33604 | **26189** |
| number of upper grids $n_u$ | **85** | 280 |
| number of lower grids $n_l$ | **214** | 459 |

The coefficient of determination (R square) of the two models, which is an index to measure the fitness of the regression model and can also be interpreted as the degree of interpretation of the model, as shown in Table 15 is calculated based on Eq. (21).

$$R^2 = 1 - \frac{SS_{res}}{SS_{tot}} = \frac{\sum_{i=1}(y_i - f_i)^2}{\sum_{i=1}(y_i - \bar{y})^2} \tag{21}$$

In Table 15, it can be found that ecept for the Nasdaq 100 index, most of the time, FNN has a better fit for the four output variables.

Table 15. Comparison of coefficient of determination between FNN and LSTM models

| Neural network type | FNN | LSTM |
| --- | --- | --- |
| **S&P 500** | | 90.490% |
| upper bound of the grid $G_{ul}$ | **97.410%** | 90.490% |
| lower bound of the grid $G_{ll}$ | **99.586%** | 97.240% |
| number of upper grids $n_u$ | 94.972% | **99.469%** |
| number of lower grids $n_l$ | **99.392%** | 94.892% |
| **Nasdaq 100** | | |
| upper bound of the grid $G_{ul}$ | 96.415% | **98.075%** |
| lower bound of the grid $G_{ll}$ | 97.088% | **99.915%** |
| number of upper grids $n_u$ | 94.942% | **99.635%** |
| number of lower grids $n_l$ | 94.737% | **98.019%** |
| **Dow Jones Industrial Average** | | |
| upper bound of the grid $G_{ul}$ | **97.704%** | 95.970% |
| lower bound of the grid $G_{ll}$ | **99.771%** | 93.429% |
| number of upper grids $n_u$ | **99.855%** | 93.685% |
| number of lower grids $n_l$ | **99.559%** | 91.285% |
| **Euro Stoxx 50** | | |
| upper bound of the grid $G_{ul}$ | **96.183%** | 94.392% |
| lower bound of the grid $G_{ll}$ | **99.324%** | 98.063% |
| number of upper grids $n_u$ | **99.297%** | 96.858% |
| number of lower grids $n_l$ | 98.405% | **99.760%** |
| **Shanghai Composite** | | |
| upper bound of the grid $G_{ul}$ | **97.157%** | 96.577% |
| lower bound of the grid $G_{ll}$ | **96.749%** | 95.545% |
| number of upper grids $n_u$ | 93.165% | **93.881%** |
| number of lower grids $n_l$ | **99.661%** | 99.003% |

Furthermore, the following four values of return of investment (ROI), maximum drawdown (MDD), volatility, and Sharpe ratio are used to examine the performance of the model in this study. In terms of methods include first buy and last sell (Buy and Sell, B&S), first sell and last buy (Sell and Buy, S&B), grid trading system robot (GTSbot) [10], Ichimoku equilibrium Figure (IK) [12], Flexible Grid trained with Fully-Connected Neural Network (FG-FNN), Flexible Grid trained with LSTM (FG-LSTM), equal-distance grid, equal-ratio grid, and flexible grid, a total of 9 methods for comparison.

First of all, the formula of ROI can refer to Eq. (22), which is the most common indicator of quantitative investment performance in investment sciences. The calculation of the ratio of investment income to cost usually presents as an annualized return to total return. The trading costs including handling fees have been included in the calculation results. The results please refer to Table 16.

$$\text{ROI} = (\text{net profit} - \text{investment costs}) \times 100\% \qquad (22)$$

Table 16. Comparison of ROI

|  | **S&P** | **Nasdaq** | **DJI** | **Euro Stoxx** | **Shanghai** |
|---|---|---|---|---|---|
| B&S | **14.392%** | 1.519% | **9.880%** | **28.340%** | -15.581% |
| S&B | -14.392% | -1.519% | -9.880% | -28.340% | 15.581% |
| GTSbot | 6.408% | 4.680% | 6.594% | 1.191% | **-0.466%** |
| IWOC | **24.111%** | **23.171%** | 7.143% | **32.087%** | -18.758% |
| FG-FNN | **11.520%** | **11.733%** | **8.849%** | 12.977% | **-3.125%** |
| FG-LSTM | 4.639% | 1.823% | 2.940% | 7.734% | -9.283% |
| Equal-distance | 5.194% | -2.972% | 3.480% | 10.748% | -4.495% |
| Equal-ratio | 4.536% | -3.625% | 3.276% | 10.764% | -4.892% |
| Flexible | 5.589% | -1.895% | 4.270% | 11.477% | -4.309% |

In terms of ROI, the FNN flexible grid achieves the top three returns in each index. In particular, the FNN flexible grid has the most ability to control the overall loss in the downward trend of prices.

Second, the comparison is the maximum drawdown, i.e., the amount of income that has fallen sharply in all periods, which is one of the indicators for evaluating investment risk. Please refer to Table 17 for the results.

Table 17. Comparison of MDD

|  | **S&P** | **Nasdaq** | **DJI** | **Euro Stoxx** | **Shanghai** |
|---|---|---|---|---|---|
| B&S | 7.960% | 12.570% | 7.290% | 4.029% | 8.529% |
| S&B | 7.491% | 10.370% | 6.580% | 7.313% | 3.103% |
| GTSbot | **0.524%** | **5.335%** | **1.000%** | **0.510%** | **1.126%** |
| IWOC | 10.345% | 12.421% | 8.606% | 5.467% | 7.640% |
| FG-FNN | 6.455% | 5.596% | 7.491% | 2.034% | 4.105% |

| FG-LSTM | 1.580% | 49.031% | 4.809% | 2.966% | 3.551% |
| Equal-distance | 5.252% | 8.594% | 4.078% | 1.513% | 4.074% |
| Equal-ratio | 5.243% | 10.096% | 4.057% | 1.529% | 4.059% |
| Flexible | 4.737% | 8.780% | 4.163% | 4.144% | 4.208% |

In terms of the maximum drawdown, it can be found that the method with a larger return on investment has a relatively higher maximum drawdown, which verifies the theory of high risk and high return in investment science. There are two reasons for the performance of the FNN flexible grid on the MDD: The risk is relatively high in the case of a relatively high return on investment. In addition, the parameters of each period are calculated by the neural network. If the parameters predicted on one period are particularly poor, it is likely to cause a large drop in a single period. Whether this MDD condition can be regarded as a single condition of outliers is yet to be verified by volatility.

In quantitative trading, risk and model stability are assessed through its volatility. The calculation method refers to Eq. (23) and the results are shown in Table 18.

$$Vol = \sigma[return] \qquad (23)$$

It can be found from Table 18 that the performance of FNN flexible grid on volatility is the top three in each index, i.e., the performance of FNN flexible grid is very stable compared with other methods. It can also be speculated that the poor performance on the MDD may be a single event, and it can be regarded as a robust investment model in most cases. The LSTM flexible grid is the best among all models in the S&P and Nasdaq data sets and is a less risky and more robust investment model.

Table 18. Comparison of volatility

|  | S&P | Nasdaq | DJI | Euro Stoxx | Shanghai |
|---|---|---|---|---|---|
| B&S | 0.06126 | 0.07671 | 0.03568 | 0.06221 | 0.01840 |
| S&B | 0.06126 | 0.07671 | 0.03568 | 0.06221 | 0.01840 |
| GTSbot | 0.01710 | 0.03589 | **0.01795** | **0.00267** | **0.00190** |
| IWOC | 0.07689 | 0.08398 | 0.04564 | 0.07337 | 0.03568 |
| FG-FNN | **0.01687** | **0.02076** | 0.02149 | **0.01730** | **0.01414** |
| FG-LSTM | **0.01672** | **0.01637** | 0.02203 | **0.01679** | 0.02030 |
| Equal-distance | 0.02450 | 0.03579 | **0.02129** | 0.01786 | 0.01590 |
| Equal-ratio | 0.02473 | 0.03629 | 0.02155 | 0.01828 | 0.01629 |
| Flexible | 0.02503 | 0.03665 | 0.02162 | 0.01862 | 0.01624 |

Finally, in terms of the Sharpe ratio, which is a model performance indicator that calculates the ratio between investment return and risk. It can also be interpreted as the reward that can be exchanged for each unit of risk. The calculation method refers to Eq. (24) and the results are as shown in Table 19.

$$Sharpe = return / \sigma [return] \qquad (24)$$

Table 19. Comparison of Sharpe ratio

|  | S&P | Nasdaq | DJI | Euro Stoxx | Shanghai |
|---|---|---|---|---|---|
| B&S | 2.349 | 0.198 | 2.769 | 4.556 | -8.466 |
| S&B | -2.349 | -0.198 | -2.769 | -4.556 | **8.466** |
| GTSbot | 3.748 | 1.304 | 3.673 | 4.466 | -2.446 |
| IWOC | 3.136 | 2.759 | 1.565 | 4.373 | -5.258 |
| FG-FNN | **6.828** | **5.651** | **4.119** | **7.499** | **-2.210** |
| FG-LSTM | 2.774 | 1.114 | 1.335 | 4.606 | -4.573 |
| Equal-distance | 2.120 | -0.830 | 1.634 | 6.017 | -2.827 |
| Equal-ratio | 1.834 | -0.999 | 1.520 | 5.887 | -3.004 |
| Flexible | 2.233 | -0.517 | 1.975 | 6.165 | -2.653 |

From Table 19, it can be found that the Sharpe ratio of FNN flexible grid is the best among all methods on the four indices. In the falling market situation, in addition to short selling, its Sharpe ratio is also the largest. From this, it can be found that the FNN flexible grid can obtain the highest reward for each unit of risk.

**5. Conclusions**

In addition to proposing a new grid trading architecture, this study effectively improves the ROI of the original model, improves the drawbacks of premature entry and exit, and utilizes SSO algorithm and ANN to assist grid trading for parameter selection and providing the model the ability to adapt to the market.

In terms of model performance, five major market indices are used as verification data, which cover the United States, Europe and China. The FNN flexible grid performs very well in Sharpe ratio. It has excellent investment return rate and model robustness, and can properly control the balance between risk and return.

This study proves that the technology combined with AI in can bring breakthroughs to the original grid trading model in quantitative trading, and can adapt to various rapidly changing market situations. In the future, looking forward to more related extended research resulting that the grid trading model can be more accurately close to human decision-making or even better than human judgment when adjusting parameters to eliminate human impulses and market emotions, and make rational market trading decisions to obtain better investment benefits.

**Acknowledgments**

This research was supported in part by the National Science and Technology Council, R.O.C (MOST 107-2221-E-007-072-MY3, MOST 110-2221-E-007-107-MY3, MOST 109-2221-E-424-002 and MOST 110-2511-H-130-002). This article was once submitted to arXiv as a temporary submission that was just for reference and did not provide the copyright.